\documentclass[12pt,a4paper]{llncs}
\usepackage{amsmath}
\usepackage{amssymb}
\usepackage{graphicx}
\usepackage{marvosym}
\usepackage{url}
\usepackage{fancyhdr}

\usepackage{graphicx}
\usepackage{fancyhdr}
\usepackage{url}
\usepackage{amsmath}
\usepackage{verbatim}  
\usepackage{amssymb}
\usepackage{eucal}
\usepackage{algpseudocode}
\usepackage{algorithm}
\usepackage{boxedminipage}
\usepackage{tabularx}
\usepackage{subfigure}
\usepackage{subcaption}

\usepackage{geometry}
\newcommand{\keywords}[1]{\par\addvspace\baselineskip
\noindent\keywordname\enspace\ignorespaces#1}

\fancypagestyle{firstpage}{\fancyhf{}
}

\begin{document}

%\mainmatter  % start of an individual contribution

\title{\LARGE{A Multiparty Commutative Hashing Protocol based on the Discrete Logarithm Problem}}

%\titlerunning{Lecture Notes in Computer Science: Authors' Instructions}

\author{\large{Daniel Zentai$^1$  \and Mihail Plesa$^2$ \and Robin Frot$^3$}}
\institute{\large{\{daniel.zentai$^1$, mihail.plesa$^2$, robin.frot$^3$\}@xtendr.io\\xtendr\\ Budapest, Hungary}}

%\author{Alfred Hofmann%
%\thanks{Please note that the LNCS Editorial assumes that all authors have used
%the western naming convention, with given names preceding surnames. This determines
%the structure of the names in the running heads and the author index.}%
%\and Ursula Barth\and Ingrid Haas\and Frank Holzwarth\and\\
%Anna Kramer\and Leonie Kunz\and Christine Rei\ss\and\\
%Nicole Sator\and Erika Siebert-Cole\and Peter Stra\ss er}

%\institute{Springer-Verlag, Computer Science Editorial,\\
%Tiergartenstr. 17, 69121 Heidelberg, Germany\\
%\url{http://www.springer.com/lncs}}


%\toctitle{Lecture Notes in Computer Science}
%\tocauthor{Authors' Instructions}

\maketitle

\thispagestyle{firstpage}

\begin{abstract}
Let $\mathcal{X}$ and $\mathcal{Y}$ be two sets and suppose that a set of participants $P=\{P_1,P_2,\dots,P_n\}$ would like to calculate the keyed hash value of some message $m\in\mathcal{X}$ known to a single participant in $P$ called the data owner. Also, suppose that each participant $P_i$ knows a secret value $x_i\in\mathcal{X}$. In this paper, we will propose a protocol that enables the participants in this setup to calculate the value $y=H(m,x_1,x_2,\dots ,x_n)$ of a hash function $H:\mathcal{X}^{n+1}\rightarrow\mathcal{Y}$ such that:
\begin{itemize}
    \item The function $H$ is a one-way function.
    \item Participants in $P\backslash\{P_i\}$ cannot obtain $x_i$.
    \item Participants other than the data owner cannot obtain $m$.
    \item The hash value $y=H(m,x_1,x_2,\dots ,x_n)$ remains the same regardless the order of the secret $x_i$ values.
\end{itemize}
\keywords{Hash functions, Discrete logarithm problem, Anonymization}
\end{abstract}

\section{Introduction}

Hash functions are very common building blocks of cryptographic protocols like digital signatures or message authentication codes. In this paper, we will propose a protocol built around the Chaum-van Heijst-Pfitzmann hash function \cite{cvhp}. The goal of our protocol is to calculate a hash value in a multiparty setup, i.e. the calculation is made by multiple participants collaboratively.\\

Our motivation was the following. Suppose we have to store personal data in a way that enables us to have access to a certain subset of non-sensitive attributes (e.g. age or height) and also enables us to keep the confidentiality of more sensitive attributes (e.g. passport number or name). Anonymization and pseudonymization may look like very similar concepts, but according to the European Union's General Data Protection Regulation (GDPR \cite{gdpr}) this is not the case. Pseudonymized data can be recovered using some secret information (e.g. a decryption key). Anonymized data on the other hand cannot be recovered under any circumstances, meaning we cannot use encryption to anonymize data.\\

In this paper, we will propose a protocol that enables a set of participants and a trusted server to calculate a hash value (i.e. anonymize their data) collaboratively.\\

We impose the following requirements regarding the process of the calculation. The function $H$ should be a one-way function, the participants (including the server) should not learn either each other's secret values or the plaintext, and the hash value should be the same regardless of which participant is the data owner (i.e. what is the order of the participants in the calculation). The reason for the last requirement is that we want to be able to detect whether two different participants have the same plaintext without knowing the plaintext itself. \\

\section{Preliminaries}
In this section, we will give some basic definitions required throughout our paper. For the interested reader, we recommend \cite{kl} for more information.\\

\begin{definition}
A function $\epsilon :\mathbb{N}\rightarrow\mathbb{R}$ is negligible if for all $c\in\mathbb{N}$ there exists an $n_0\in\mathbb{N}$ such that for all $n\geq n_0$ we have $\epsilon(n)<n^{-c}.$
\end{definition}

\begin{definition}
A hash function $h$ is a deterministic polynomial time algorithm with arbitrary input length and fixed output length.\\
\end{definition}

\begin{definition}
A hash function $h$ is said to be one-way if it is
\begin{itemize}
    \item Easy to compute, i.e. $h$ outputs $h(x)$ in polynomial time for all $x\in D(h)$.
    \item Hard to invert, i.e. given a hash value $y$ no probabilistic polynomial time algorithm can calculate an $x$ for which $y=h(x)$ with non-negligible probability.
\end{itemize}
\end{definition}

\begin{definition}
A hash function is said to be collision resistant if no probabilistic polynomial time algorithm can succeed in finding two values $x\neq x'\in D(h)$ such that $h(x)=h(x')$ with non-negligible probability.
\end{definition}

Throughout this paper, we will assume that the discrete logarithm problem defined below is hard, i.e. it cannot be solved by any probabilistic polynomial time algorithm.

\begin{definition}
If $G$ is a cyclic group and $a,b\in G$ are group elements then the discrete logarithm problem includes finding an $x\in G$ such that $a^x=b$.
\end{definition}

There are several cryptosystems based on the hardness of the discrete logarithm problem. The most popular ones include the Diffie-Hellman key exchange \cite{dh} and the ElGamal encryption \cite{eg}.

\section{Related Work}
There are several anonymization techniques available, including but not limited to $k$-anonymity, differential privacy, or synthetic data \cite{anon}. These methods are designed either to add noise to the data, add some artificial data to the real data set, or generalize the data such that the data subject is no longer identifiable from it.\\

Our method is a bit different in a few aspects. First of all, we propose a purely cryptographic method, the protocol requires the collaboration of multiple participants, and the anonymized (i.e. hashed) version of a fixed data record remains the same regardless of which participant owns the plaintext. The main advantage of this approach is that this way we allow the participants to calculate the intersection of their data set without revealing sensitive identifiers.\\

To ensure that the hash of a fixed message $m$ remains the same regardless of which participant owns the plaintext, we have to use commutative hashing. In \cite{comhash} the authors propose $h(x,y)=f(\min(x,y),\max(x,y))$ where $f$ is a collision-resistant hash function. Unfortunately, a similar approach won't work in our case since the participants know nothing about each other's secret values.

\begin{definition}
    A hash function $h:\mathcal{X}^n\rightarrow\mathcal{Y}^n$ is commutative, if $h(x_1,x_2,\dots,x_n)=h(\sigma(x_1,x_2,\dots,x_n))$ for all $x_1,x_2,\dots,x_n$ and all $\sigma$ permutations.
\end{definition}

Note that we cannot ensure collision resistance if we use a commutative hash function, since $(x_1,x_2,\dots,x_n)\neq\sigma(x_1,x_2,\dots,x_n)$ (unless $\sigma$ is the identity permutation), but $h(x_1,x_2,\dots,x_n)=h(\sigma(x_1,x_2,\dots,x_n))$ by definition. Therefore the most we can do without losing the desired functionality is to ensure one-way property.

\section{Hashing Protocol}

As a building block of our protocol, we will use the discrete logarithm-based collision-resistant hash function defined by Chaum, van Heijst, and Pfitzmann \cite{cvhp}
\\
\begin{definition}
Let $p$ be a prime number such that $q=\frac{p-1}{2}$ is also a prime number, and let $a$ and $b$ be primitive elements of the multiplicative group of $GF(p)$. Let $h:GF(q)\times GF(q)\rightarrow GF(p)\backslash\{0\}$ be the following function: $$h(x,y)=a^x\cdot b^y \mod p$$
\end{definition}

The hash function $h$ defined above is collision-resistant if the discrete logarithm problem is hard \cite{cvhp}.\\

Now, suppose that we have a set of participants $P=\{P_1, P_2,\dots, P_n\}$ and a trusted server represented by a special participant $S\notin P$. Also, suppose that the data owner (i.e. the only participant who knows $m$ is $P_1$, therefore $P_1$ initiates the hashing protocol).\\

Let $h$ be the Chaum-van Heijst-Pfitzmann hash function and $\Pi=(Gen, Enc, Dec)$ a public key encryption scheme. Suppose that each participant $P_i$ has two randomly generated keys $x_i,y_i\in GF(q)$. Also, suppose that $P_1$ wants to hash the message $m\in GF(q)$. In addition, we will suppose that all the messages in the protocol are sent over a secure channel.\\

Our protocol proceeds as follows:\\

\noindent\begin{boxedminipage}{\textwidth}
\begin{center}
$\ $\\
\textbf{Multiparty Commutative Hashing Protocol}\\
\end{center}
\begin{enumerate}
\item Upon receiving an upload request from $P_1$, $S$ generates $n$ random numbers $r_1,r_2,\dots,r_n$ and sends them to the participants in $P$.

\item Upon receiving $r_1$ from $S$, $P_1$ calculates $h_1=h(x_1+m,y_1)$ and sends $$h_1| Enc_{K^{pub}_S}(r_1)$$ to $S$, where $K^{pub}_S$ is the public key of $S$.

\item For all $i=2,\dots, n$, upon receiving $r_i$ from $S$, $P_i$ calculates $h_i=h(x_i,y_i)$ and sends $$h_i| Enc_{K^{pub}_S}(r_i)$$ to $S$.

\item Upon receiving $h_i| Enc_{K^{pub}_S}(r_i)$ from $P_i$ for all $i=1,\dots, n$, $S$ checks the random values $r_1,r_2,\dots, r_n$ and stores $\prod_{j=1}^nh_j$ if these are the same values $S$ sent out at the first step. Otherwise, $S$ outputs an error message.
 \\
\end{enumerate}
\end{boxedminipage}
\\
Note that the purpose of the $r_i$ values is not authentication, these just make it possible for the server to check that every participant took part in the calculation and therefore $\prod_{j=1}^nh_j$ can be stored as the output of the protocol.\\

\begin{theorem}
If the discrete logarithm problem is hard, then $H(x_1,\dots,x_n,y_1,\dots,y_n)=\prod_{i=1}^nh(x_i,y_i)$ is a one-way function.
\end{theorem}
\textit{Proof.} Recall that if the discrete logarithm problem is hard, then $h$ is a collision-resistant hash function. Suppose that $H$ is not a one-way function, thus there exists a PPT algorithm $\mathcal{A}$ that can successfully find a corresponding input to a given value of $H$. We will use $\mathcal{A}$ to find a collision in $h$.\\

From the construction of $H$, it immediately follows that 
\begin{equation}
\begin{aligned}
H(x_1,\dots,x_n,y_1,\dots,y_n) & = \prod_{i=1}^nh(x_i,y_i)\\
& =\prod_{i=1}^na^{x_i}\cdot b^{y_i}\\
& =a^{\sum_{i=1}^nx_i}\cdot b^{\sum_{i=1}^ny_i}\\
& =h\left(\sum_{i=1}^nx_i,\sum_{i=1}^ny_i\right)
\end{aligned}
\end{equation}

Now let $\alpha =H(x,0,\dots,0,y,0,\dots,0)$ and run algorithm $\mathcal{A}$ on $\alpha$. Suppose that $\mathcal{A}(\alpha)=x_1^*,\dots,x_n^*,y_1^*,\dots,y_n^*$.\\

From this, we can successfully find a preimage of $h(x,y)$ since 

\begin{equation}
\begin{aligned}
h(x,y) & = H(x,0,\dots,0,y,0,\dots,0)\\
& = H(x_1^*,\dots,x_n^*,y_1^*,\dots,y_n^*)\\
& = h\left(\sum_{i=1}^nx_i^*,\sum_{i=1}^ny_i^*\right)
\end{aligned}
\end{equation}

And that is a contradiction since $h$ is collision-resistant (therefore also a one-way function) if the discrete logarithm problem is hard.
\begin{flushright}
$\square$
\end{flushright}
Note that $H$ is still collision-resistant in a somewhat weaker sense. If we fix the $x_i$ and $y_i$ values (e.g. we do not let the participants to change their keys) then it is hard to find two messages $m\neq m'$ with the same hash value.

\section{Hashing with elliptic curves}

Although the Chaum-van Heijst-Pfitzmann hash is a provable secure construction, it is not widely used in practice because of its running time. One way to make a discrete logarithm-based protocol more applicable in practice, is to use elliptic curves.

\begin{definition}
Let $p$ be a prime number and $\mathcal{E}_p$ be an elliptic curve over the finite field $GF(p)$. Also, let $P,Q$ be points of $\mathcal{E}_p$. The elliptic curve discrete logarithm problem includes finding a $k\in\mathbb{Z}$ such that $k\cdot P=Q$.
\end{definition}

The elliptic curve discrete logarithm problem is thought to be even harder than the aforementioned classical discrete logarithm problem. The Chaum-van Heijst-Pfitzmann hash function can be defined as follows using elliptic curves.

\begin{definition}
Let $p$ be a prime number and $\mathcal{E}_p$ be an elliptic curve over the finite field $GF(p)$, moreover assume that $|\mathcal{E}_p|$ is a prime $n$. Let $A$ and $B$ be base points of $\mathcal{E}_p$ such that the discrete logarithm of $B$ is not known. Let $h:(\mathbb{Z}\backslash n\mathbb{Z})^2\rightarrow \mathcal{E}_p\backslash\{\mathcal{O}\}$ be the following function: $$h(k,l)=k\cdot A+l\cdot B$$
\end{definition}

\begin{theorem}
The hash function $h$ defined above is collision-resistant if the elliptic curve discrete logarithm problem is hard.    
\end{theorem}
\textit{Proof.} Suppose that we can find a collision efficiently, i.e. two pairs of integers $(k,l)\neq(k',l')$ such that $$h(k,l)=k\cdot A+l\cdot B=k'\cdot A+l'\cdot B=h(k',l').$$
We can assume that $l'\neq l$. Otherwise, we would have $kA=k'A$ and thus $k=k'$.
Since $n$ is prime, $l-l'$ is invertible modulo $n$ and this gives $$\frac{k-k'}{l'-l}\cdot A=B.$$
We successfully solved the elliptic curve discrete logarithm problem.
\begin{flushright}
$\square$
\end{flushright}

Again, suppose that we have a set of participants $P=\{P_1, P_2,\dots, P_n\}$ and a trusted server represented by a special participant $S\notin P$. Also, suppose that the data owner (i.e. the only participant who knows $m$ is $P_1$, therefore $P_1$ initiates the hashing protocol.\\

Let $h$ be the Chaum-van Heijst-Pfitzmann hash function and $\Pi=(Gen, Enc, Dec)$ a public key encryption scheme. Suppose that each participant $P_i$ has two randomly generated keys $k_i,l_i\in \mathbb{Z}$. Also, suppose that $P_1$ wants to hash the message $m$ where $m$ is a point of an elliptic curve $\mathcal{E}_p$ over the finite field $GF(p)$

Now our modified protocol proceeds as follows:\\

\noindent\begin{boxedminipage}{\textwidth}
\begin{center}
$\ $\\
\textbf{Multiparty Anonymization Protocol with Elliptic Curves}\\
\end{center}
\begin{enumerate}
\item Upon receiving an upload request from $P_1$, $S$ generates $n$ random numbers $r_1,r_2,\dots,r_n$ and sends them to the participants in $P$.

\item Upon receiving $r_1$ from $S$, $P_1$ calculates $h_1=h(k_1+m,l_1)$ and sends $$h_1| Enc_{K^{pub}_S}(r_1)$$ to to $S$, where $K^{pub}_S$ is the public key of $S$.

\item For all $i=2,\dots, n$, upon receiving $r_i$ from $S$, $P_i$ calculates $h_i=h(k_i,l_i)$ and sends $$h_i| Enc_{K^{pub}_S}(r_i)$$ to $S$.

\item Upon receiving $h_i| Enc_{K^{pub}_S}(r_i)$ from $P_i$ for all $i=1,\dots, n$, $S$ checks the random values $r_1,r_2,\dots, r_n$ and stores $\sum_{j=1}^nh_j$ if these are the same values $S$ sent out at the first step. Otherwise, $S$ outputs an error message.
 \\
\end{enumerate}
\end{boxedminipage}

\section{Implementation results}

We instantiated the protocol both over the multiplicative group of integers modulo $p$ and over the elliptic curve SECP256k1 \cite{chen2023recommendations}. The purpose of our experiment was to analyze the running time with respect to the number of participants. We discovered that the elliptic curve implementation is $10 \times$ faster than the implementation over the multiplicative group of integers. For each number of participants, we run $100$ trials of the protocol and average the running time. The results are presented in Table \ref{table:1}. We used the Levenberg-Marquardt algorithm to determine the coefficients of a linear polynomial that best fit the data \cite{ranganathan2004levenberg}. Let $N$ be the number of participants. We denote by $T_{\mathbb{Z}_p}\left(N\right)$ and $T_{\mathcal{E}_p}\left(N\right)$ the running time in seconds of the protocol implemented over $\mathbb{Z}_p$, respectively $\mathcal{E}_p$. The discovered polynomials are shown in (4) and (5):

\begin{equation}
    T_{\mathcal{E}_p}\left(N\right) = 0.008N - 0.733 
\end{equation}
\begin{equation}
    T_{\mathbb{Z}_p}\left(N\right) = 0.13N - 42.06
\end{equation}

\begin{table}[ht!]
\centering
\begin{tabularx}{0.8\textwidth} { 
  | >{\raggedright\arraybackslash}X 
  | >{\centering\arraybackslash}X 
  | >{\raggedleft\arraybackslash}X | }
 \hline
 \textbf{Num. of participants} & \textbf{Running time over $\mathcal{E}_p$ (s)} & \textbf{Running time over $\mathbb{Z}_p$ (s)}\\
 \hline
 4  & 0.028  & 0.280 \\
 \hline
 8  & 0.058 & 0.561  \\
 \hline
 16  & 0.111 & 1.122  \\
 \hline
 32  & 0.222 & 2.245  \\
 \hline
 64  & 0.442 & 4.511  \\
 \hline
 128  & 0.882 & 9.065  \\
 \hline
 256  & 1.772 & 18.318 \\
 \hline
 512  & 3.551 & 37.178  \\
 \hline
 1024  & 7.135 & 76.174  \\
 \hline
 2048  & 14.431 & 160.599  \\
 \hline
 4096  & 29.474 & 352.496  \\
 \hline
 8192  & 61.157 & 837.490  \\
 \hline
 16384  & 132.225 & 2242.677  \\
\hline

\end{tabularx}
\caption{The running time (s) with respect to the number of participants}
\label{table:1}
\end{table}

\begin{figure}[h]
\centering
\includegraphics[width=.8\textwidth]{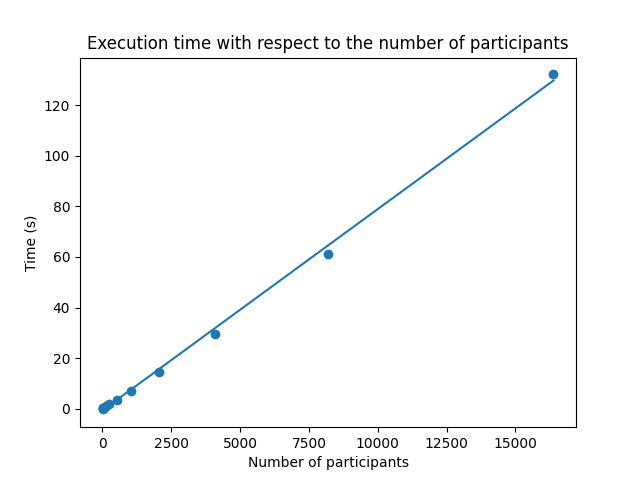}
\vspace{-5pt}
\caption{The running time of the protocol implemented over $\mathcal{E}_p$} 
\label{fig1}
\end{figure}

\begin{figure}[h]
\centering
\includegraphics[width=.8\textwidth]{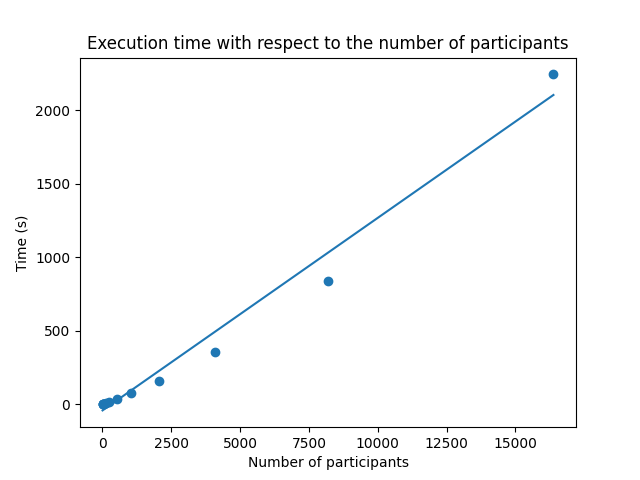}
\vspace{-5pt}
\caption{The running time of the protocol implemented over $\mathbb{Z}_p$} 
\label{fig2}
\end{figure}

Figures \ref{fig1} and \ref{fig2} show the running time with respect to the number of participants as well as the polynomial found by the curve fitting algorithm when implemented over $\mathcal{E}_p$ and over $\mathbb{Z}_p$.

All experiments were performed in Python on an Apple M1 Max platform. 

\newpage
\section{A threshold protocol}
One possible relaxation of the protocol is to ensure that any $k$-element subset of the participants can calculate the anonymous ID for some fixed $1<k\leq n=|P|$. As a building block, we can use Shamir's secret sharing scheme \cite{sh} which works as follows.\\
Let $\mathbb{F}$ be a finite field and let $s_0\in\mathbb{F}$ be the secret we want to share with the participants. Let $a_1,\dots,a_{k-1}$ be elements of $\mathbb{F}$ chosen uniformly at random and $f(x)=s_0 + \sum_{i=1}^{k-1}a_ix^i$. Each participant $P_i$ is given a point $(x_i,f(x_i))$ of the polynomial $f$. With Lagrange interpolation, any $k$ element subset of the participants can now calculate the secret $s_0$ as follows.
$$s_0=f(0)=\sum_{i=1}^kf(x_i)\cdot\ell_i$$
where $$\ell_i=\prod_{j=1,j\neq i}^k\frac{x_j}{x_j-x_i}$$ 
Now we can use this secret sharing scheme to modify our anonymization protocol in a way that any $k$ element subset of $P$ can calculate the anonymous IDs.\\
Let $h$ be the Chaum-van Heijst-Pfitzmann hash function and $\Pi=(Gen, Enc, Dec)$ a Fully Homomorphic Encryption (FHE) scheme. Suppose that each participant $P_i$ has a randomly generated key $x_i\in GF(q)$. Also, suppose that $P_1$ wants to hash the message $m\in GF(q)$. In addition, we will suppose that all the messages in the protocol are sent over a secure channel.\\

\noindent\begin{boxedminipage}{\textwidth}
\begin{center}
$\ $\\
\textbf{Multiparty Threshold Anonymization Protocol}\\
\end{center}
\begin{enumerate}
\item First, $S$ generates two secret $s_0$ and $t_0$ and distributes them using Shamir's secret sharing scheme to the members in $P$. $S$ uses two different polynomials $f(x)=s_0 + \sum_{i=1}^{k-1}a_ix^i$ and $g(x)=t_0 + \sum_{i=1}^{k-1}b_ix^i$. Let $x_1,\dots,x_n\in GF(p)$ secret random elements generated by the respective $P_i$.

\item Upon receiving $Enc_{K^{pub}_i}(x_i)$ from $P_i$, $S$ sends $f(Enc_{K^{pub}_i}(x_i))=Enc_{K^{pub}_i}(f(x_i))$ and $g(Enc_{K^{pub}_i}(x_i))=Enc_{K^{pub}_i}(g(x_i))$ to $P_i$ where $K^{pub}_i$ is the public key of $P_i$.

\item Every pair $(P_i,P_{i+1})$ shares with $S$ the quotient $x_{i+1}/x_i$ with the \textbf{Multiply} protocol described below.

\item Upon receiving an anonymization request from $P_1$, $S$ randomly chooses a $k$ element subset $Q\subseteq P$. Without loss of generality suppose now that $Q=\{P_1,\dots, P_k\}$. Additionally, $S$ generates $k$ random values $r_1,r_2\dots ,r_k$ and sends $r_i$ to $P_i\in Q$.

\item For every participants $P_i\in Q$, $S$ calculates $\ell_i=\prod_{j=1,j\neq i}^k\frac{x_j}{x_j-x_i}$ with the stored quotients and sends $\ell_i$ to $P_i$.

\item Upon receiving $r_1$ and $\ell_1$ from $S$, $P_1$ calculates $h_1=h(m+f(x_1)\cdot\ell_1,g(x_1)\cdot\ell_1)$ and sends $$h_1|Enc_{K_S^{pub}}(r_1)$$ to $S$ where $K_S^{pub}$ is the public key of $S$.

\item For all $i=2,\dots, k$, upon receiving $r_i$ from $S$, $P_i$ calculates $h_i=h(f(x_i)\cdot\ell_i,g(x_i)\cdot\ell_i)$ and sends $$h_i| Enc_{K^{pub}_S}(r_i)$$ to $S$.

\item Upon receiving $h_i| Enc_{K^{pub}_S}(r_i)$ from $P_i$ for all $i=1,\dots, k$, $S$ stores 
$$\prod_{i=1}^kh_i=a^{m+\sum_{i=1}^kf(x_i)\cdot\ell_i}\cdot b^{\sum_{i=1}^kg(x_i)\cdot\ell_i}=a^{m+s_0}\cdot b^{t_0}$$ 
if the random values $r_1,r_2,\dots, r_k$ are the same values $S$ sent out at the first step. Otherwise, $S$ outputs an error message.
\\
\end{enumerate}
\end{boxedminipage}
\\

Note that in step $5$, the server $S$ needs to calculate $\frac{x_i}{x_i-x_j}$, where $x_j$ is not known by $P_i$ since it is a secret value generated by $P_j$. We will use a subroutine that solves this problem. The subroutine uses the fact that $\frac{x_i}{x_i-x_j}$ can be calculated securely if we can calculate the product $x_k^{-1}\cdot x_{k+1}$ securely, since $$\frac{x_i}{x_i-x_j}=\left(\frac{x_i-x_j}{x_i}\right)^{-1}=\left(1-\frac{x_j}{x_i}\right)^{-1}=\left(1-x_i^{-1}\cdot x_j\right)^{-1}$$
and, if $j>i$, $$\frac{x_j}{x_i} = \prod_{k=i}^{j-1}\frac{x_{k+1}}{x_k}.$$

\noindent\begin{boxedminipage}{\textwidth}
\begin{center}
$\ $\\
\textbf{Two-Party Multiplication Protocol}\\
\end{center}
\begin{enumerate}
\item \textbf{Multiply($x,y$)} takes two inputs from two different parties: $x$ from $P_1$ and $y$ from  $P_2$. 
\item $P_1$ generates a random number $r_1$ and sends $r_1\cdot x$ to $P_2$ who then generates a random number $r_2$, calculates $r_1\cdot x\cdot r_2\cdot y$ and sends it to the server. 
\item The server also generates a random value $r_S$, calculates $r_S\cdot r_1\cdot x\cdot r_2\cdot y$ and sends it back to $P_1$.
\item $P_1$ calculates $r^{-1}_1\cdot r_S\cdot r_1\cdot x\cdot r_2\cdot y=r_S\cdot x\cdot r_2\cdot y$ and sends it to $P_2$.
\item $P_2$ calculates $r^{-1}_2\cdot r_S\cdot x\cdot r_2\cdot y=r_S\cdot x\cdot y$ and sends it to the server.
\item The server calculates $r^{-1}_S\cdot r_S\cdot x\cdot y=x\cdot y$.
 \\
\end{enumerate}
\end{boxedminipage}
\\

Note that in both versions of the protocol, if the server is corrupted, it can brute-force the hash values. Indeed, if $P_i$ is the data owner during the first run then $S$ learns $h(x_j,y_j)$ (or already knows $(s_0,t_0)$ in the threshold protocol) for all $j\in P\backslash\{P_i\}$. And if $P_u\neq P_i$ is the data owner during the second run then $S$ learns $h(x_i,y_j)$ (or $h(f(x_i),g(x_i)$ in the threshold protocol) as well. We can prevent this if the participants agree on a random number $R$ using some group key exchange protocol and run the protocol on the message $R\cdot m$ instead of $m$. Other participants besides $S$ cannot perform this attack since every message is sent over a secure channel.

Also note that it is possible to double the size of the plaintext space if $P_1$ sends $h(m_1+x_1,m_2+y_1)$ (or $h(m_1+f(x_1)\cdot \ell_1,m_2+g(x_1)\cdot \ell_1)$ in the threshold case) with $(m_1,m_2)\in GF(p)^2$.

%\newpage

\section{Conclusion}
We proposed a protocol that enables a set of participants to calculate a one-way commutative hash function collaboratively in a way that all of the participants have to take part in the computation, and this fact is verifiable by a trusted server and no participant other than the data owner has access to the plaintext. We also proposed a threshold protocol where any $k$-element subset of the participants can run the protocol successfully.\\

A possible future direction would be to construct a decentralized version of the protocol or make the protocol post-quantum.\\

Our implementation was purely experimental and was not made for efficiency. One possible direction of research is to create an efficient implementation of the protocol using low-level parallelizable instruction or FPGA \cite{brown1996fpga}\cite{peleg1996mmx}.

\end{document}